\documentclass[sigconf]{acmart}
\usepackage{amsmath,amsfonts}
\usepackage{algorithm,algorithmic}
\usepackage{multirow,multicol}
\usepackage{graphicx}
\usepackage{textcomp}
\usepackage{xcolor}
\usepackage{booktabs}
\usepackage{comment}
\usepackage{footmisc}
\usepackage{mathrsfs}
\usepackage{enumitem}
\usepackage{array}
\usepackage{float}
\usepackage{hyperref}
\usepackage{units}
\usepackage{subfigure} 
\AtBeginDocument{%
  \providecommand\BibTeX{{%
    \normalfont B\kern-0.5em{\scshape i\kern-0.25em b}\kern-0.8em\TeX}}}

\copyrightyear{2022}
\acmYear{2022}
\setcopyright{acmcopyright}
\acmConference[WWW '22]{Proceedings of the ACM Web Conference 2022}{April 25--29, 2022}{Virtual Event, Lyon, France}
\acmBooktitle{Proceedings of the ACM Web Conference 2022 (WWW '22), April 25--29, 2022, Virtual Event, Lyon, France}
\acmPrice{15.00}
\acmDOI{10.1145/3485447.3512071}
\acmISBN{978-1-4503-9096-5/22/04}

\begin{document}
\title{AutoField: Automating Feature Selection in Deep \\Recommender Systems}

\author{Yejing Wang$^{1}$, Xiangyu Zhao$^{1*}$, Tong Xu$^2$, Xian Wu$^3$}
\thanks{* Xiangyu Zhao is corresponding author.}
\affiliation{
	\institution{$^1$City University of Hong Kong, $^2$University of Science and Technology of China, $^3$ Tencent Inc}
	\country{}
}
\email{adave631@gmail.com, xianzhao@cityu.edu.hk, tongxu@ustc.edu.cn, kevinxwu@tencent.com}
\renewcommand{\shortauthors}{Yejing Wang, et al.}

\begin{abstract}
Feature quality has an impactful effect on recommendation performance. Thereby, feature selection is a critical process in developing deep learning-based recommender systems.
Most existing deep recommender systems, however, focus on designing sophisticated neural networks, while neglecting the feature selection process.
Typically, they just feed all possible features into their proposed deep architectures, or select important features manually by human experts.
The former leads to non-trivial embedding parameters and extra inference time, while the latter requires plenty of expert knowledge and human labor effort.
In this work, we propose an AutoML framework that can adaptively select the essential feature fields in an automatic manner.
Specifically, we first design a differentiable controller network, which is capable of automatically adjusting the probability of selecting a particular feature field; 
then, only selected feature fields are utilized to retrain the deep recommendation model.
Extensive experiments on three benchmark datasets demonstrate the effectiveness of our framework. 
We conduct further experiments to investigate its properties, including the transferability, key components, and parameter sensitivity. 
\end{abstract}

\keywords{Feature Selection, Recommender System, AutoML}

\ccsdesc[500]{Information systems~Recommender systems}

\maketitle

\section{Introduction} \label{sec:introduction}

\begin{figure}
    \centering
    \includegraphics[width=\linewidth]{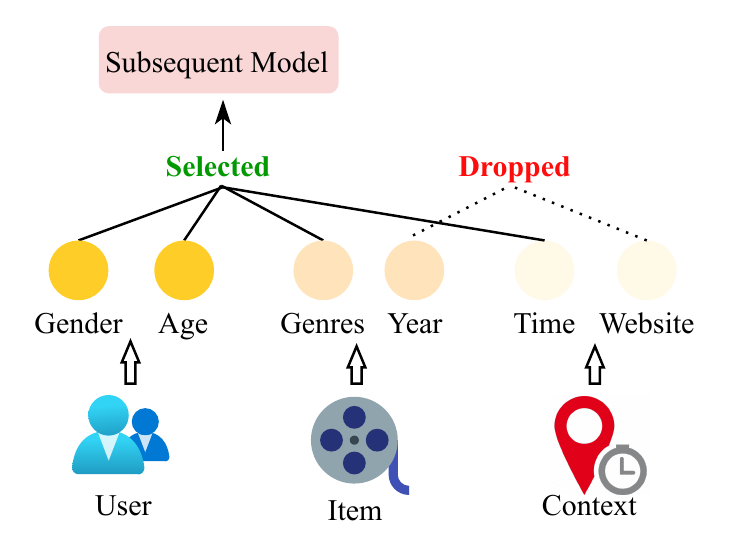}
    \vspace{-10mm}
    \caption{An example of feature selection process} 
    \label{fig:feature}
\end{figure}

In the world-wide-web era, customers are overwhelmed with overloading information, making it difficult to retrieve valuable information. 
Well-designed recommender systems, which significantly alleviate this issue by presenting selected items that best match user preference~\cite{schafer2007collaborative,rendle2010factorization}, are widely employed in a variety of scenarios, including shopping websites~\cite{qu2016product,wang2017deep}, online movie/music platforms~\cite{zhao2016user,zhao2017social}, and location-based services~\cite{liu2017experimental}. 

With the rapid growth of deep learning theories and techniques, deep learning-based recommender systems could capture user preference accurately with their strong feature representation and inference capacities~\cite{zhang2019deep}.
The majority of existing works spend lots of effort on designing novel and sophisticated neural architectures; however, feature selection, the process of selecting a subset of informative and predictive feature fields, has received little attention~\cite{mitchell1997machine,friedman2017elements,bengio2013representation}.
They typically feed their proposed deep recommender systems with all possible features without dropping the unavailing ones~\cite{he2017neural,he2017neuralcf,guo2017deepfm}.
This often necessitates additional computations to learn all the feature embeddings, thereby further slowing down the model inference. 
Sometimes, ineffective features can even be detrimental to model performance~\cite{nadler2005prediction}.

\begin{figure*}[ht]
    \centering
    \includegraphics[width=0.9\linewidth]{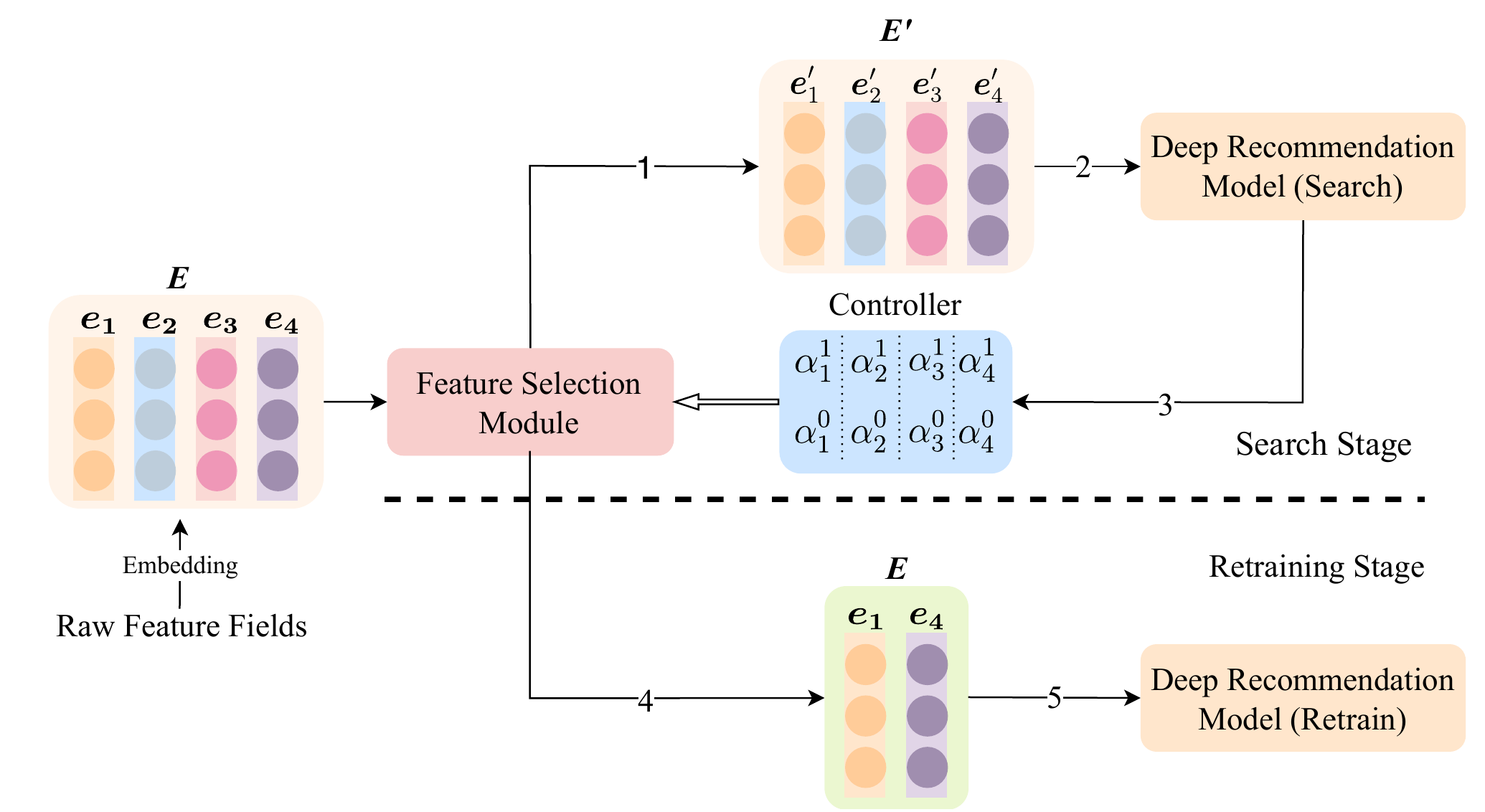}
    \caption{Framework overview. Step 1: Soft feature selection; Step 2: Update the recommendation model; Step 3: Update the controller; Step4: Hard feature selection with well-trained controller; Step 5: Retrain the recommendation model. }
    \label{fig:framework}
\end{figure*}

\vspace{1mm}
Several conventional feature selection methods can be leveraged in developing deep recommender systems. 
First, hand-crafted feature selection by human experts is generally used in traditional recommender systems. In fact, this is a trial-and-error method manually conducted by human experts, which requires necessary expert knowledge and laborious human efforts.
In addition, grid search~\citep{hinton2012practical,glorot2010understanding} or random search~\citep{bergstra2012random} algorithms, which exhaustively or randomly sample the solutions from all possible feature selection options, are comparably accurate but prohibitively expensive in computations and time.
Statistics and traditional machine learning techniques can also be utilized in feature selections, such as filter methods via correlation analysis~\citep{hall1999correlation,yu2003feature}, as well as embedded methods via Lasso regularization~\citep{fonti2017feature} and decision trees~\citep{ke2017lightgbm}.
However, these methods usually fail in deep recommender systems. For instance, filter methods neglect the dependency between feature selection and downstream deep recommendation models; Embedded methods are sensitive to the recommendation models' strong structural assumptions.
With recent advances in automated machine learning (AutoML)~\citep{yao2018taking}, there is a tremendous interest in developing deep recommender systems via AutoML technologies. However, most efforts aim to select predictive feature 
interactions~\citep{liu2020autofis,luo2019autocross,song2019autoint}, while few of them focus on selecting predictive features directly.

\vspace{1mm}
In this paper, we aim to propose an efficient framework that automates the feature selection process when building deep recommender systems. This task is rather challenging due to the huge search space of all feature selection options, i.e., $2^N$ with $N$ the number of feature fields. To address this issue, we develop an AutoML based framework AutoField, which can automatically identify an optimal subset of predictive feature fields from the whole search space. In addition, AutoField is capable of processing discrete categorical features and simultaneously evaluating the contributions of all feature fields in a parallel fashion. The experiment results on three benchmark datasets demonstrate the effectiveness of our proposed model. 
We summarized our major contributions as follows:
\begin{itemize}[leftmargin=*] 
    \item We propose a novel AutoML-based framework, AutoField, which can select predictive features automatically, improving the model performance and inference efficiency;
    \vspace{1mm}
    \item We design a controller architecture to dynamically moderate the probability pairs deciding whether to select or drop feature fields. Decisions are instructive for other deep recommender systems on the same dataset;
    \item We carry out extensive experiments on real-world benchmark datasets with investigating the effectiveness, efficiency, and transferability of the proposed framework. 
\end{itemize}

We organize the rest of this paper as follows. In Section \ref{sec:framework}, we present the main modules of the proposed framework, the optimization strategy, and the retraining method. Section \ref{sec:EXP} details the experiments, including the research questions we care about and corresponding experimental results. In Section \ref{sec:Relat}, we introduce related works in brief. Finally, we conclude this work in Section \ref{Conclu}.
\section{The Proposed Framework} \label{sec:framework}
In this section, we introduce the technical details of the proposed AutoField framework. We first provide an overview of AutoField. And then, we detail the important modules, the AutoML-based optimization algorithm, and the retraining method. 

\subsection{Framework Overview}
In this subsection, we will provide an overview of the AutoField framework. AutoField aims to select the optimal subset of feature fields from the entire feature space automatically. 
The framework overview is illustrated in Figure \ref{fig:framework}. Briefly, the proposed framework consists of two stages, the search stage, and the retraining stage. To find the optimal subset of feature fields, we first update the framework parameters in the search stage.  Then, we select the best feature fields based on learned parameters and feed them into subsequent deep recommender systems in the retraining stage.

In the search stage, we first initialize the framework parameters. Then, we feed all the feature fields to the feature selection module, which bridges the controller with the deep recommendation model by selecting feature fields in a differentiable manner according to the controller parameters. After that, the deep recommendation model predicts the user preference. With prediction errors on the training set, we update the parameters of the deep recommendation model. Simultaneously, the controller is optimized based on validation data. At the end of this stage, we obtain the well-trained parameters in the controller for the next stage.

All possible feature fields are fed into the deep recommendation model during the search stage, but only a part of the feature fields will be selected for the final deep recommendation model. As a result, we retrain the model after the search stage. Specifically, the feature selection module first selects the predictive feature fields based on the controller parameters. To be noted, the behavior of feature selection module is different in two stages. We conduct hard selection in the retraining stage but use soft selection in the search stage. Subsequently, with selected feature fields, the deep recommendation model is adapted and retrained.

\subsection{Deep Recommendation Model}\label{subsec:utilize}

\begin{figure}[t]
	\centering
	\includegraphics[width=\linewidth]{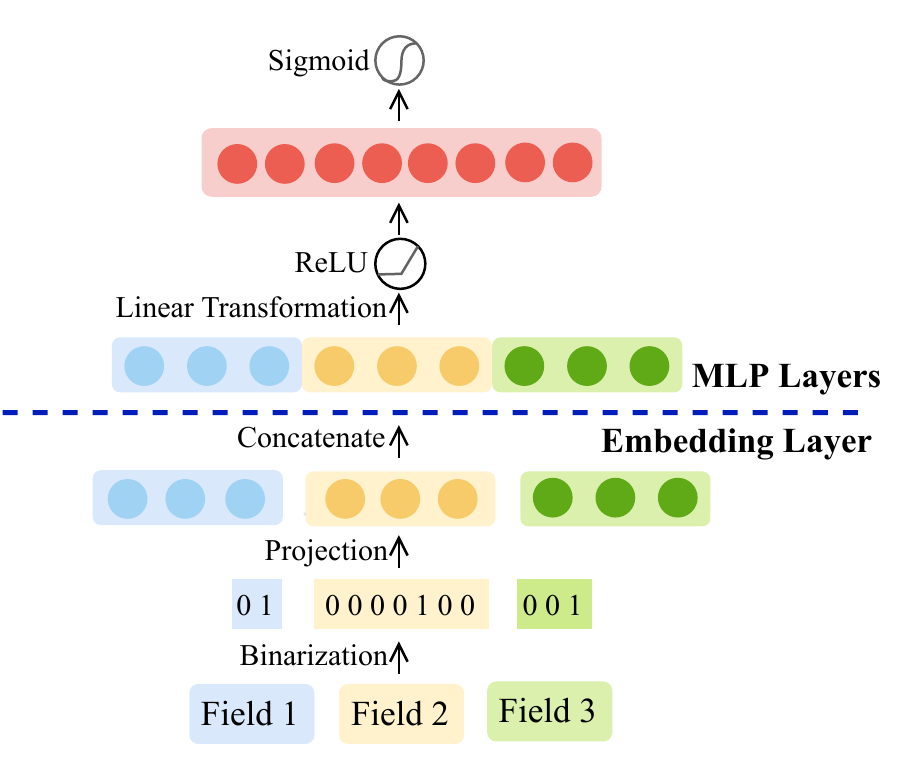}
	\caption{Architecture of deep recommendation model}
	\label{fig:my_label}
\end{figure}

We will introduce the architecture of the deep recommendation model in this subsection.
As visualized in Figure \ref{fig:my_label}, a typical deep recommendation model has two essential components: the embedding layer and the MLP layer.

\subsubsection{\textbf{Embedding Layer}}

The embedding layer is a basic component widely used in deep recommender systems, which converts the categorical inputs into low-dimensional continuous vectors via two steps, i.e., binarization and projection.

\textbf{Binarization.} Categorical inputs are usually transformed into binary vectors. 
The dimension of these vectors is equivalent to the number of unique feature values in the corresponding feature fields. For example, the feature field ``Gender" has two unique feature values, ``Male" and ``Female", then they can be represented by the two-dimensional binary vectors $[1,0]$ and $[0,1]$. 

\textbf{Projection.} Then, given ${N}$ input feature fields of a user-item interaction data, the $N$ features can be represented as the concatenation of their binary 
vectors, i.e., $ \boldsymbol{x} =[\boldsymbol{x}_1,\boldsymbol{x}_2,\cdots,\boldsymbol{x}_N]$, $\boldsymbol{x}_n \in \mathbb{R}^{D_n}$, where ${\boldsymbol{x}_n}$ is the binary vector for the ${n^{th}}$ feature field, and ${D_n}$ is the corresponding dimension.

Due to the high sparsity and varied length, binary vectors still require further processing in the projection step. For $\displaystyle \boldsymbol{x}_n, \forall n\in[1,N]$, we project it into low-dimensional space as:
\begin{gather}
{\boldsymbol{e}_n = \boldsymbol{A}_n\boldsymbol{x}_n}
\end{gather}
where {${\boldsymbol{A}_n\in \mathbb{R}^{d\times D_n}}$ } is a learnable weight matrix, and ${d}$ is the pre-defined embedding size of the projection space. Then, the final embedding of the user-item interaction data is:
\begin{gather}
{\boldsymbol{E} = [\boldsymbol{e}_1,\boldsymbol{e}_2,\cdots,\boldsymbol{e}_N}]\label{equ:Eout}
\end{gather}

\subsubsection{\textbf{MLP Layer}} Multi-layer Perceptron (MLP) structure is a common component in deep recommender systems~\cite{qu2016product,guo2017deepfm}. It mines information from feature embedding by linear transformations and non-linear activations, formulated as:
\begin{gather}
	\boldsymbol{h}_0 = \boldsymbol{E} \\
	\boldsymbol{h}_{m+1} = \mathrm{ReLU}(\boldsymbol{W}_m\boldsymbol{h}_m+\boldsymbol{b}_m), 0 \leq m \leq M \label{equ:hidden}\\
	\boldsymbol{\hat y} = \sigma (\boldsymbol{W}_{M+1}\boldsymbol{h}_{M+1} + {b}_{M+1}) \label{equ:outlay}
\end{gather}
where the initial input of MLP, say $\boldsymbol{h}_0$, is the output of the embedding layer.
For Equation (\ref{equ:hidden}), {$\boldsymbol{W}_m,\boldsymbol{b}_m,\boldsymbol{h}_m$ } respectively stand for the weight matrix, bias vector, and output of ${m}^{th}$ layer, where ${M}$ is the number of MLP layers before the output layer. 
For Equation (\ref{equ:outlay}), $\boldsymbol{W}_{M+1},\boldsymbol{b}_{M+1}$ are the weight matrix and bias vector of the output layer, and ${\sigma(\cdot)}$ is the activation function depending on specific recommendation tasks.

\subsection{Controller}\label{subsec:controller}
\begin{figure}
    \centering
    \includegraphics[width=0.8\linewidth]{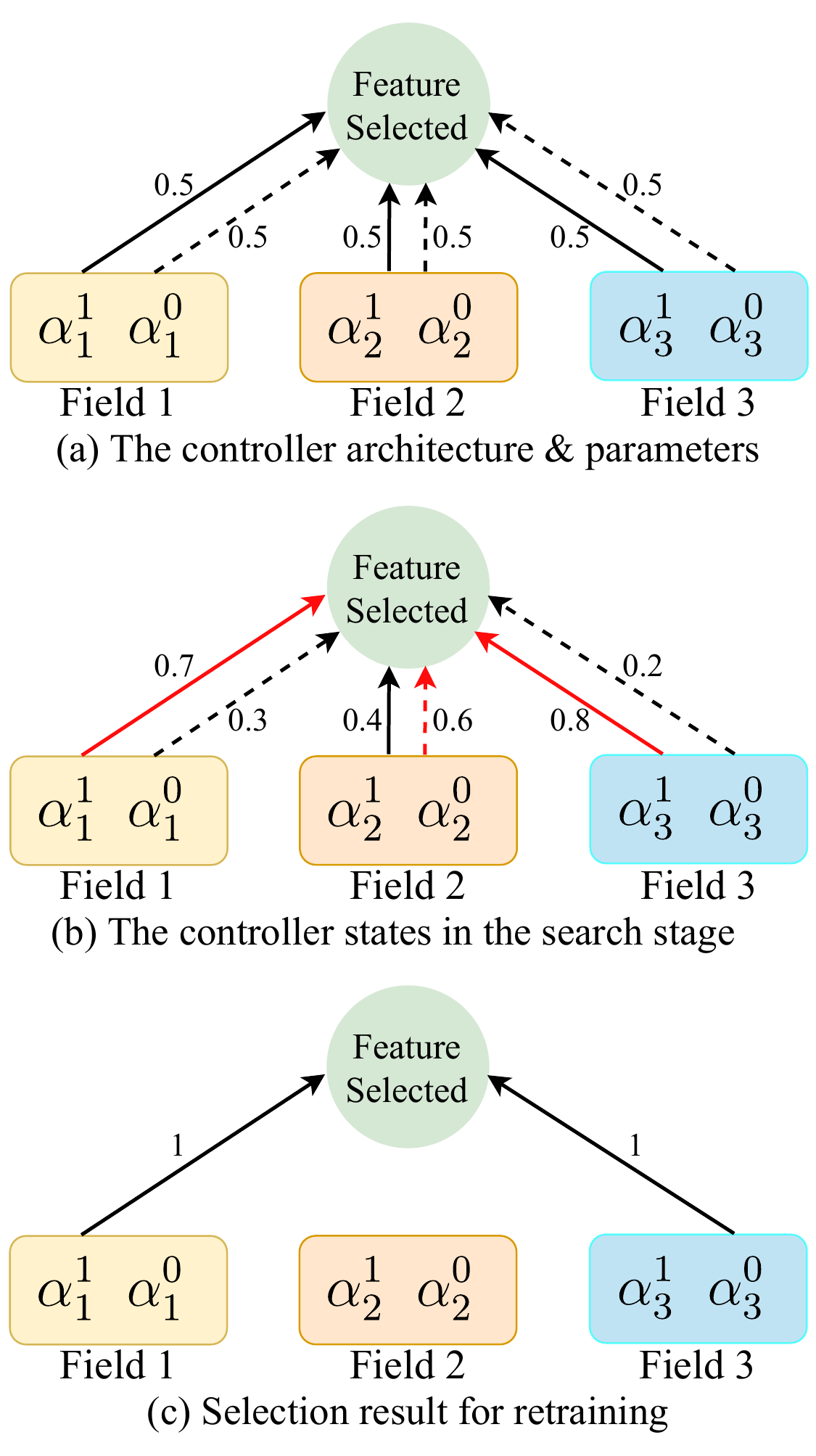}
    \caption{The selection process of controller}
    \label{fig:controller}
\end{figure}

In this subsection, we will introduce our feature selection's search space and the controller structure.
For an input dataset with $N$ feature fields, we have two choices for each feature field, i.e., ``selected'' or ``dropped'', thus the size of the whole search space is $2^N$. With this huge space, common search space construction methods like encoding ~\citep{pham2018efficient} are incompatible with the practical demand for efficiency. Motivated by~\cite{liu2018darts}, we define our search space as a directed acyclic graph in Figure~\ref{fig:controller}. 
To be more specific, we utilize $N$ parallel nodes to select $N$ feature fields respectively, where the $n^{th}$ node is a 2-dimensional vector containing two parameters $(\alpha_n^1, \alpha_n^0)$. In other words, there are $2\times N$ parameters in total for the $N$ feature fields, which control the behaviors of the feature selection process. We denote $\alpha_n^1$ as the probability of selecting a feature field and $\alpha_n^0$ as that of dropping the feature field, thus we have $\alpha_n^1 + \alpha_n^0 = 1$.

As visualized in Figure~\ref{fig:controller} (a), we first assign a pair of equivalent $(\alpha_n^1, \alpha_n^0)$ to each feature field, i.e., $\alpha_n^1=\alpha_n^0=0.5$. During the training progress as in Figure~\ref{fig:controller} (b), the parameter $\alpha_n^1$ of predictive features (e.g., Field 1 and 3) will increase, while $\alpha_n^0$ will decrease. For non-predictive features (e.g., Field 2), the situation is the opposite. Finally, as in Figure~\ref{fig:controller} (c), the feature fields with higher probability $\alpha_n^1$ will be selected for model retraining, while the ones with higher $\alpha_n^0$ will be dropped.

\subsection{Feature Selection Module}\label{subsec:selection}
In this subsection, we will further detail the feature selection process of AutoField in the search stage.

According to the controller's parameter pair ${(\alpha_n^1,\alpha_n^0)}$, each feature field will be selected or dropped for the final recommendation model. To be specific, we could define the feature selection process as the expectation of sampling, formulated as:
\begin{gather}
    \boldsymbol{e}_n' =  (\alpha_n^1 \cdot \boldsymbol{1}  + \alpha_n^0 \cdot \boldsymbol{0}  ) \cdot \boldsymbol{e}_n = \alpha_n^1\boldsymbol{e_n} \label{weisum}\\
 \boldsymbol{E}' = {[\boldsymbol{e}_1',\boldsymbol{e}_2',\cdots,\boldsymbol{e}_N']}
 \label{equ:EoutWei}
\end{gather}
\noindent where $\boldsymbol{e}_n$ is the embedding of ${n}^{th}$ feature. $\boldsymbol{1}$ and $\boldsymbol{0}$ are all-one and all-zero vectors with the same length of $\boldsymbol{e}_n$. $\boldsymbol{e}_n'$ is the soft selection of the ${n}^{th}$ feature field in a weighted sum embedding manner. Accordingly, $\boldsymbol{E}'$ is the soft selection version of all feature embeddings $\boldsymbol{E}$ of Equation (\ref{equ:Eout}).

This soft selection manner, however, cannot completely eliminate the influence of suboptimal feature fields on the final recommendation, leading to a gap between the search and retraining stages.
Therefore, a hard selection method is highly desirable. 
To address this issue, we introduce the Gumbel-Max trick~\citep{gumbel1954statistical} to simulate the hard selection process based on the controller's parameters ${(\alpha_n^1,\alpha_n^0)}$. To be specific, we can draw a hard selection $z_n$ as: 
\begin{gather}
z_n=\mathrm{one\_hot}\left(  \mathrm{arg max}\left[\log \alpha^0_n+g_0,\log \alpha^1_n+g_1\right] \right) \label{equ:gumbelp}\\
\text { where } g_j = -\log \left(-\log \left(u_j\right)\right)\nonumber\\
u_j \sim \mathrm{Uniform}(0,1) \quad \forall j \in [0,1]\nonumber
\end{gather}
where ${\{g_j\}}$ are independent and identically distributed (i.i.d) gumbel noises. For ${n}^{th}$ feature field, if we define the selection behavior as ${z_n\boldsymbol{e_n}}$, then its expectation is equivalent to the weighted sum version as in Equation (\ref{weisum}).

Note that ${z_n}$ in Equation (\ref{equ:gumbelp}) is obtained from the non-differentiable operation, i.e., $\mathrm{one\_hot(argmax(\cdot))}$. To apply the gradient optimization strategy, we approximate it by the softmax operation \citep{jang2016categorical} to make the architecture differentiable as:
\begin{gather}
p^j_n=\frac{\exp \left(\left(\log \alpha^j_{n}+g_{j}\right) / \tau\right)}{\exp \left(\left(\log \alpha^1_n+g_1\right) / \tau\right)+\exp \left(\left(\log \alpha^0_n+g_0\right) / \tau\right)}
\end{gather}
where ${p^j_n}$ is the new probability for selecting ($j=1$) or dropping ($j=0$) the ${n^{th}}$ feature, and $\tau$ it a hyperparameter called temperature. 
Since ${[\log \alpha^0_n+g_0,\log \alpha^1_n+g_1]}$ in Equation (\ref{equ:gumbelp}) is a pair of extreme values, i.e., 
${p_n^i}$ is very close to $1$ or $0$, then we can use ${p_n^i}$ to simulate the hard selection, which bridges the aforementioned gap between model search and retraining stages.

Finally, the feature selection in Equation (\ref{weisum}) can be rewritten as below:
\begin{gather}
\boldsymbol{e}_n' =  (p_n^1 \cdot \boldsymbol{1}  + p_n^0 \cdot \boldsymbol{0}  ) \cdot \boldsymbol{e}_n = p_n^1\boldsymbol{e_n} \label{equ:gumexpect}
\end{gather}

Based on Equation (\ref{equ:gumexpect}), we can obtain the $\boldsymbol{E}'$ as in Equation (\ref{equ:EoutWei}), and then replace the embeddings $\boldsymbol{E}$ of Equation (\ref{equ:Eout}) by $\boldsymbol{E}'$, so as to conduct feature selection with the proposed controller.

\subsection{Optimization}
In the above subsections, we introduce the architecture of the deep recommendation model, the controller, and formulate the feature selection process based on controller parameters. Next, we will detail the optimization.

In the search stage, there are two sets of parameters to be optimized, i.e., those of the deep recommendation model and the controller. 
We denote the parameters of the deep recommendation model as $\boldsymbol{W}$ and the parameters of the controller as $\mathcal{A}$. $\boldsymbol{W}$ and $\mathcal{A}$ should not be jointly learned from the same batch of training data as traditional supervised learning~\citep{pham2018efficient}. This is because the learning process of $\boldsymbol{W}$ and $\mathcal{A}$ are highly interdependent, which would lead to a severe over-fitting problem if we update both of them on the same training batch.

Inspired by DARTS~\citep{liu2018darts}, we formulate a bi-level optimization problem~\citep{anandalingam1992hierarchical} to tackle the over-fitting issue by updating $\mathcal{A}$ and $\boldsymbol{W}$ alternatively based on different sets of data:
\begin{equation}
\begin{aligned}
&\min _{\mathcal{A}} \mathcal{L}_{v a l}\left(\boldsymbol{W^{*}}(\mathcal{A}), \mathcal{A}\right) \label{equ:optimization}\\
&\text { s.t. }\boldsymbol{ W^{*}}(\mathcal{A})=\arg \min _{\boldsymbol{W}} \mathcal{L}_{\text {train }}\left(\boldsymbol{W}, \boldsymbol{\mathcal{A}}^{*}\right)
\end{aligned}
\end{equation}
where $\mathcal{L}_{\text {train }}$ and $\mathcal{L}_{v a l}$ are the binary cross-entropy loss (BCE)\footnote{We use BCE here since all tasks in our experiments are binary classification problems. It is straightforward to use other loss functions according to different recommendation tasks.} $(\boldsymbol{y},\boldsymbol{\hat y}) = \boldsymbol{y} \cdot \log \boldsymbol{\hat y}+\left(1-\boldsymbol{y}\right) \cdot \log \left(1-\boldsymbol{\hat y}\right)$ on the training set and validation set, respectively. $\boldsymbol{y}$ are the ground truth, and $\boldsymbol{\hat y}$ are the model predictions.

\begin{algorithm}[t]
	\caption{Optimization Algorithm of AutoField Framework}\label{Alg:OPT}
	\raggedright
	{\bf Input}: Raw feature fields $\{\boldsymbol{e}_n\}$, number of selected feature fields $K$, controller update frequency $f$\\
	{\bf Output}: $K$ selected feature fields\\
	\begin{algorithmic} [1] 
		\STATE $t = 0$ ($t$ records the training steps) \label{Line:set}
		\WHILE{not converge} \label{Line:While}
		\STATE Sample a mini-batch from the training dataset
		\STATE Update $\boldsymbol{W}$ according to the constraint in Equation (\ref{equ:optimization}) 
		\IF{$t\;\%\;f =0$}
		\STATE Sample a mini-batch from the validation dataset
		\STATE Update $\mathcal{A}$ according to the Equation (\ref{equ:optimization}) \label{Line:TrainA}
		\ENDIF
		\STATE $t = t + 1$
		\ENDWHILE
	\end{algorithmic}
\end{algorithm}

The detailed optimization of the AutoField framework is illustrated in Algorithm \ref{Alg:OPT}. We manually set hyper-parameters: (i) the number of finally selected feature field ${K}$; and (ii) the frequency ${f}$ of updating controller parameters $\mathcal{A}$, compared with that of $\boldsymbol{W}$. In other words, we train $\mathcal{A}$ once after training $\boldsymbol{W}$ of ${f}$ times. To be specific, we first set a flag ${t}$ to record the training steps (line 1).  Then, we train $\boldsymbol{W}$ and $\mathcal{A}$ to converge (line 2), where we train the recommendation model parameters $\boldsymbol{W}$ with every mini-batch of training data (line 3-4) and train the controller parameters with a mini-batch of validation data every ${f}$ time steps (line 5-8). 

\subsection{Retraining Stage}\label{subsec:retraining}
In the search stage, we feed all raw features into the deep recommendation model. Thus, suboptimal feature fields might hurt the model performance. To address this issue, we need to retrain the final deep recommendation model with only selected feature fields.
In this subsection, we will introduce how to select feature fields according to the well-trained controller, and retrain the framework.

After optimization, we have well-learned controller parameters ${\{(\alpha^1_1,\alpha^0_1),\cdots (\alpha^1_N,\alpha^0_N)\}}$. Then, we can select ${K}$ feature fields with highest ${\alpha^1_n}$ for retraining the deep recommendation model, where ${K}$ is manually predefined.

In Section \ref{subsec:utilize}, we design the embedding layer and MLP layer with all raw feature fields. In the retraining stage, we should adapt the embedding layer and the MLP layer for only ${K}$ selected feature fields. To be specific, we should omit the embedding parameters of dropped feature fields and adjust the input size of the first MLP layer, from ${N \times d}$ to ${K \times d}$. With the selected $K$ predictive feature fields and corresponding new architecture, we can then retrain the deep recommendation model via back-propagation on the training dataset.

It is noteworthy that the selected $K$ predictive feature fields are not restricted to the basic deep recommendation model in Section \ref{subsec:utilize}. In other words, AutoField's selected feature fields have outstanding transferability to other deep recommendation models. We will demonstrate this claim in Section \ref{subsec:TransEXP}.

\section{Experiment}\label{sec:EXP}
In this section, we conduct extensive experiments to evaluate the effectiveness of the proposed framework. Specifically, the main research questions we care about are as follows: 
\begin{itemize}[leftmargin=*]
\item \textbf{RQ1:} How does AutoField perform compared with other feature selection methods? 
\item \textbf{RQ2:} Are the selected feature fields of AutoField transferable? 
\item \textbf{RQ3:} What is the impact of the components in AutoField?   
\item \textbf{RQ4:} What is the influence of key hyper-parameters? 
\item \textbf{RQ5:} Does AutoField really select out the optimal subset of feature fields?  
\end{itemize}

\begin{table*}[!ht]
\caption{Overall performance}
\label{tab:Overall Performance}
\begin{tabular}{@{}c|ccc|ccc@{}}
\toprule[1pt]
\multirow{2}{*}{Selection   Method} & \multicolumn{3}{c|}{Criteo}                                          & \multicolumn{3}{c}{Avazu}                                                               \\
                                    & Dropped Feature Fields                 & AUC $\uparrow$             & Logloss $\downarrow$        & Dropped Feature Fields                                   & AUC  $\uparrow$           & Logloss  $\downarrow$       \\ \midrule
All Fields                          & None                          & 0.8027          & 0.4491          & None                                          & 0.7769          & 0.3816          \\
PCA                                 & {[}8,  10, 16,  20, 28, 33{]}    & 0.7999          & 0.4516          & {[}0, 1, 2, 5, 8, 12, 13, 15, 16, 18, 21{]}  & 0.7751          & 0.3829          \\
LASSO                               & {[}9, 17, 18, 20, 21{]}              & 0.8000          & 0.4514          & {[}0, 1, 2, 3, 6, 9, 13, 14, 16, 17, 19{]}        & 0.7583          & 0.3919          \\
GBR                                 & {[}13, 17, 20{]}                   & 0.8022          & 0.4497          & {[}0, 1, 2, 5, 8, 12, 13, 15, 16, 18, 21{]}         & 0.7751          & 0.3829          \\
GBDT                                & {[}7, 12,  16,  19, 21, 23{]} & 0.8010          & 0.4506          & {[}0, 1, 7, 9, 10, 11, 12, 13, 14, 15, 20{]}          & 0.7467          & 0.3967          \\
IDARTS                              & Unstable                         & 0.7985          & 0.4518          & Unstable                                            & 0.7554          & 0.3920          \\
\textbf{AutoField}                  & \textbf{{[}13, 14, 17, 20, 21, 33{]}} & \textbf{0.8029} & \textbf{0.4490} & \textbf{{[}1, 5, 7, 8, 12, 13, 15, 16, 18, 19, 21{]}} & \textbf{0.7773} & \textbf{0.3813} \\ \bottomrule[1pt]
\end{tabular}
\end{table*}

\subsection{Datasets}\label{subsec:EXPdata}
We evaluate the overall performance of the AutoField framework on two benchmark datasets: 
\begin{itemize}[leftmargin=*]
\item\textbf{Avazu\footnote{https://www.kaggle.com/c/avazu-ctr-prediction/}:} This dataset was provided for a Kaggle click-through rates (CTR) prediction competition. There are 40 million user-click records in this dataset with 22 feature fields. Parts of them are anonymous. 
\item\textbf{Criteo\footnote{https://www.kaggle.com/c/criteo-display-ad-challenge/}:} This is a real-world industry benchmark dataset for predicting the CTR of online advertisements. It includes 45 million user-click records on delivered ads with 26 categorical feature fields and 13 numerical feature fields. We convert the numerical feature fields to categorical\footnote{https://www.csie.ntu.edu.tw/ r01922136/kaggle-2014-criteo.pdf}. All feature fields are anonymous.
\end{itemize}

For both datasets, we use $80\%$ for training, $10\%$ for
validation, and the rest $10\%$ for testing.  

\subsection{Evaluation metric}
We use the AUC score (Area Under the ROC Curve) and Logloss to evaluate recommendation performance. It is noteworthy that a higher AUC score and lower Logloss at 0.001-level could indicate a significant improvement in click-through rates prediction tasks~\citep{guo2017deepfm}. Besides, to compare the efficiency of models, we record the inference time per batch and the total retraining time of models with selected feature fields.

\subsection{Implementation}
In this subsection, we detail the implementation of AutoField framework. 
For the embedding layer, we set the embedding size of all feature fields as 16. 
For the MLP layer, we use two fully-connected layers of size $[16,8]$ and use the activation function $\mathrm{ReLU}$. 
Since the click-through rates prediction task could be regarded as a binary classification problem, we use the activation function $\mathrm{Sigmoid}$ in the output layer. For the controller, as introduced in Section \ref{subsec:controller}, we assign two parameters $(\alpha_n^1, \alpha_n^0)$ to $n^{th}$ feature field and apply $\mathrm{Gumbel{-}Softmax}$ to simulate hard selection. For other parameters, we set all learning rates as 0.0001, the batch size as 2048, and the drop-out rate as $0.2$. The temperature in Equation (\ref{equ:gumbelp}) is $ \tau = \max(0.01, 1-0.00005\cdot t)$, where $t$ is the training step. 

Note that our implementation is based on a public Pytorch library for recommendation models\footnote{https://github.com/rixwew/pytorch-fm}. Feature selection baselines are included in the Python library sckit-learn~\cite{scikit-learn}. All experiments are conducted on a GPU of GEFORCE RTX 3090.

\subsection{Overall Performance(RQ1)}\label{subsec:OAPer}
To evaluate the effectiveness of AutoField, we conduct extensive experiments with different feature selection methods based on the deep recommendation model mentioned in Section \ref{subsec:utilize}. The representative baselines include PCA~\citep{wold1987principal}, LASSO~\citep{tibshirani1996regression}, GBR~\citep{breiman1997arcing}, GBDT~\citep{breiman1997arcing} and IDARTS~\citep{jiang2019improved}. Details of baseline settings can be found in Appendix \ref{ap:baselineIntro}. AutoField drops 11 feature fields for Avazu ($50\%$) and 6 feature fields for Criteo ($15\%$). For a fair comparison, baselines also drop the same number of feature fields (or less). The overall performance is shown in Table \ref{tab:Overall Performance}. We can find that: 

\begin{itemize}[leftmargin=*]
\item For both datasets, applying AutoField to select predictive features could enhance the recommendation performance. It is noteworthy that we dropped $50\%$ of the feature fields in Avazu, but achieve better performance. 

\item PCA and LASSO can select some informative feature fields like AutoField. 
However, they miss some other predictive feature fields and select some trivial feature fields. 
The main reason is that PCA neglects the correlations between ground truth $\boldsymbol{y}$ and feature fields, while LASSO is sensitive to the strong structural assumptions of the deep recommendation model. 
\item For gradient boosting methods, i.e., GBR and GBDT, they obtain a better selection result on Criteo than the two previous baselines. However, GBDT gets an improper selection on Avazu. 
The performance of gradient boosting methods, which assign importance weights to feature fields, is highly dependent on the predicting ability of origin classifiers. The original GBDT method (without MLP layer) only achieves an AUC of $0.52$ on Avazu, which indicates its ineffectiveness as an original classifier. 
Besides, the effectiveness of gradient boosting methods lies in the ensemble structures, which require lots of computations. 
\item Another AutoML-based method IDARTS cannot generate stable selection results, so we show its average performance of various selection results in Table \ref{tab:Overall Performance}. The reason is that IDARTS considers all feature fields in a single search space and applies softmax over all of them, which might be significantly disturbed by the stochastic training process and stuck at the local optimum. AutoField introduces Gumbel-Softmax to prevent local optimum. 
\item Besides, with selected features, the retraining time of the deep recommendation model can be reduced significantly, such as $50\%$ on the Avazu dataset.
\end{itemize}

In general, compared with state-of-the-art feature selection methods, AutoField can select more predictive feature fields that further improve model recommendation performance.

\subsection{Transferability Analysis(RQ2)}\label{subsec:TransEXP}

\begin{table}[]
\caption{Transferability analysis on Avazu}
\label{tab:EXPtransfer}
\begin{tabular}{@{}c|cccc@{}}
\toprule[1pt]
Model                       & AUC              & Logloss          & Infer./ms & Time Saved \\ \midrule
\multirow{2}{*}{FM}         & 0.7763          & 0.3817          & 27.6           &                 \\
                            & \textbf{0.7771*} & \textbf{0.3813*} & \textbf{26.4*}   & 3.98  \%   \\ \midrule
\multirow{2}{*}{DeepFM}     & 0.7799          & 0.3796          & 46.2            &                 \\
                            & \textbf{0.7818*} & \textbf{0.3785*} & \textbf{43.8*}   & 4.97  \%    \\ \midrule
\multirow{2}{*}{xDeepFM}    & 0.7846         & 0.3771          & 85.8         &                 \\
                            & \textbf{0.7852*} & \textbf{0.3767*} & \textbf{81.0*}   & 5.10   \%   \\ \midrule
\multirow{2}{*}{IPNN}       & 0.7836        & 0.3782          & 58.8            &                 \\
                            & \textbf{0.7837} & \textbf{0.3777} & \textbf{52.2*}   & 10.98   \%   \\ \midrule
\multirow{2}{*}{Wide\&Deep} & 0.7776          & 0.3810          & 46.8          &                 \\
                            & \textbf{0.7781} & \textbf{0.3808} & \textbf{39.0*}   & 17.02  \%  \\ \midrule
\multirow{2}{*}{DCN}        & 0.7797          & 0.3798          & 58.2         &                 \\
                            & \textbf{0.7799} & \textbf{0.3797} & \textbf{54.0*}   & 7.30   \%   \\ \bottomrule[1pt]
\end{tabular}
\\Inference is based on a batch of test data (batchsize=2048).
\\``\textbf{{\Large *}}'' indicates the statistically significant improvements (i.e., two-sided t-test with $p<0.05$) over the original model.
\end{table}
In this subsection, we will investigate the transferability of AutoField's feature selection results. We apply its selected feature fields to six advanced deep recommendation models, namely Factorization Machine (FM)~\citep{rendle2010factorization}, DeepFM~\citep{guo2017deepfm}, xDeepFM~\citep{lian2018xdeepfm}, IPNN~\citep{qu2016product}, Wide\&Deep (WD)~\citep{cheng2016wide}, DeepCrossNet (DCN)~\citep{wang2017deep} on the Avazu dataset. In Table \ref{tab:EXPtransfer}, the first line of each model is the performance with all possible feature fields, while the second line is that with only selected feature fields by AutoField. The AUC score, Logloss, and average inference time on a test batch are compared. It can be observed that: 
\begin{itemize}[leftmargin=*]
\item For all models, applying AutoField's selection results not only improves the model performance but also reduces the inference time, which can enhance the online inference efficiency. This shows that the AutoField framework makes a general feature selection, which can also benefit other recommender systems.  
\item By applying AutoField's selection results, we can observe that the recommendation performance improves more on the first three FM-based models, while the inference time reduces more on other models like IPNN, xDeepFM, Wide\&Deep, DCN. 
\end{itemize}

In summary, based on experimental results, we can conclude that the feature selection results of AutoField have a strong transferability to other deep recommendation models.

\begin{table}[ht]
\caption{Ablation study on Avazu}
\label{tab:ALres}
\begin{tabular}{@{}cccc@{}}
\toprule[1pt]
Model              & AUC             & Logloss         & Dropped Features Fields     \\ \midrule
AL1                & 0.7769          & 0.3816          & None \\
AL2                & 0.7615          & 0.3883          & Not Stable   \\
AL3                & 0.7554          & 0.3920          & Not Stable   \\ 
\textbf{AutoField} & \textbf{0.7773} & \textbf{0.3813} & \textbf{{[}1,5,7,8,12,13,15,16,18,19,21{]}}            \\ \bottomrule[1pt]
\end{tabular}
\end{table}

\subsection{Ablation Study (RQ3)}
In this subsection, we will study the contributions of important components of AutoField. 
We mainly analyze two components and design the following three variants:
\begin{itemize}[leftmargin=*] 
\item \textbf{AL-1}: In this variant, we use the $\mathrm{argmax}$ operation to make an automated selection rather than selecting ${K}$ feature fields with the highest scores. In other words, AL-1 simply selects all feature fields with $\alpha_n^1 > 0.5, \forall n \in [1,N]$.
\item \textbf{AL-2}: This variant conducts the feature selection using original $\mathrm{Softmax}$ instead of Gumbel-Softmax.
\item \textbf{AL-3}: This variant involves both the above modifications in AL-1 and AL-2.
\end{itemize}

The results are shown in Table \ref{tab:ALres}. We can observe that: 
\begin{itemize}[leftmargin=*] 
\item Only the proposed AutoField framework can generate a stable result of feature selection. 
\item Making decisions by only $\mathrm{argmax}$ cannot achieve the goal of the feature selections. For example, AL-1 selects all the feature fields. The reason is that this variant overlooks the comparison with other feature fields.  
\item Original $\mathrm{Softmax}$ can lead to suboptimal model performance, such as AL-2. The reason might be that soft selection results in a gap between the search stage and the retraining stage. 
\end{itemize}

With the ablation study above, we further validated the contribution of vital components of  AutoField.

\begin{figure}[ht]
 \centering
 {\subfigure{\includegraphics[width=0.49\linewidth]{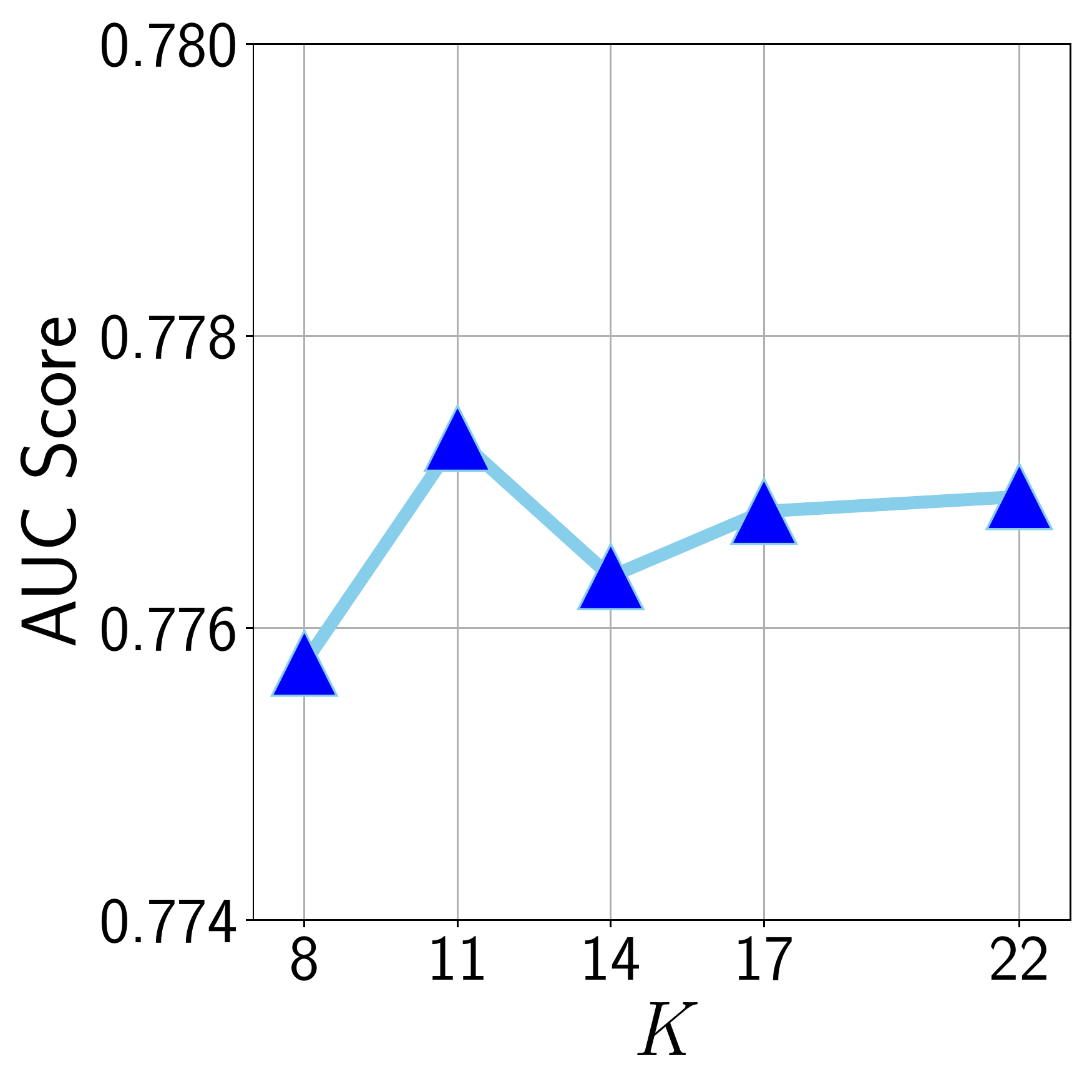}}}
 {\subfigure{\includegraphics[width=0.49\linewidth]{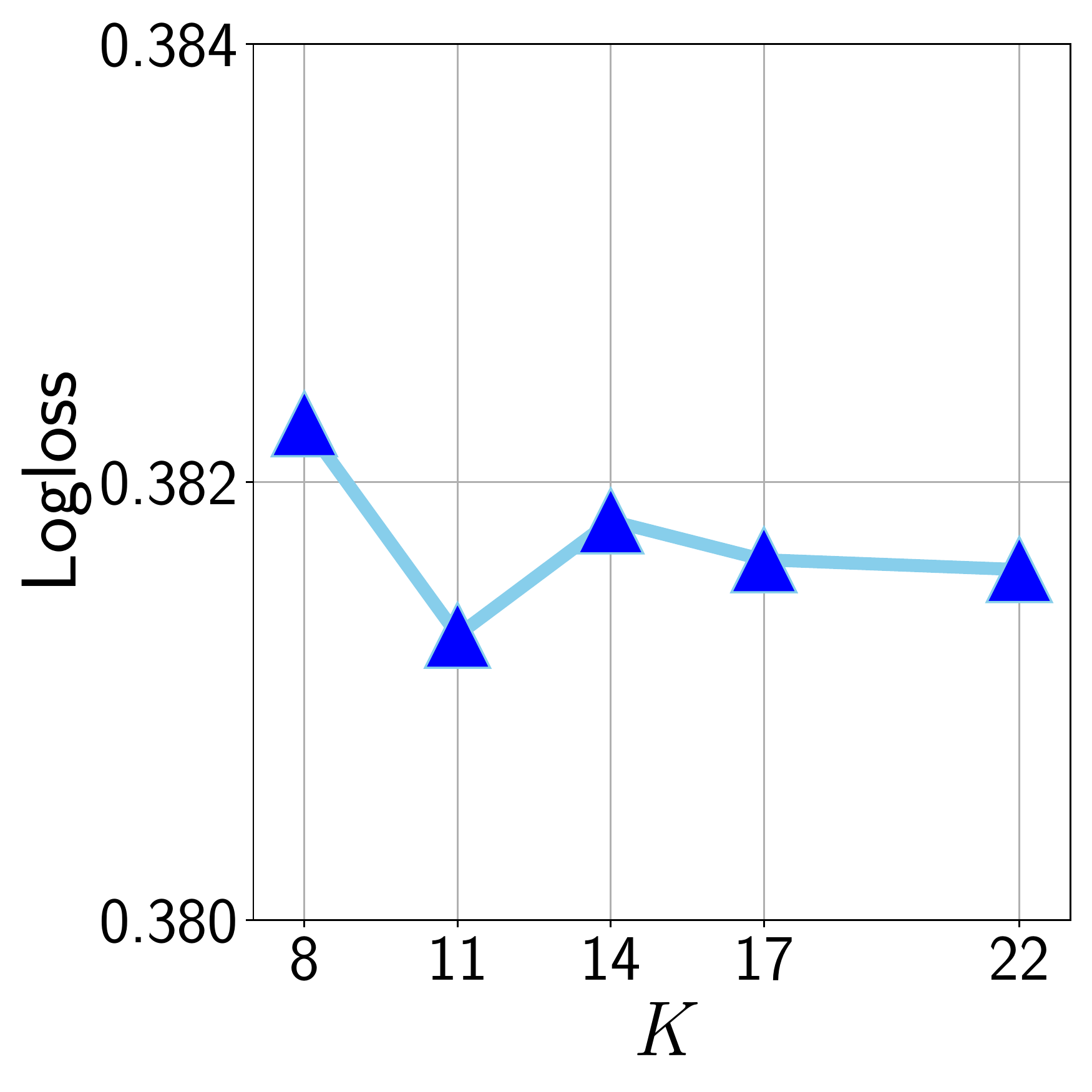}}}
 \caption{Parameter analysis on Avazu}
 \label{fig:param}
\end{figure}

\subsection{Parameter Analysis (RQ4)}
In this subsection, we aim to investigate the importance of hyper-parameters. The most crucial one for AutoField is the number of selected feature fields, i.e., ${K}$. We set ${K}$ as $[8,11,14,17]$ and fix all other hyper-parameters, then select the optimal feature fields via AutoField on Avazu dataset. The results are visualized in Figure \ref{fig:param}. We can find that:
\begin{itemize}[leftmargin=*] 
\item The optimal ${K}$ is 11, i.e., $K^*=11$.
\item With larger ${K}$ like $K=14$, AutoField will introduce some trivial features that hurt the model performance. When ${K}$ becomes even larger like $K=17$, AutoField could select feature fields that perform close to ${K=22}$, i.e., the baseline \textit{All Fields} in Table \ref{tab:Overall Performance}. 
\item When the number of feature fields is less than $K^*$ like $K=8$, the model performance degrades significantly. The reason is that AutoField will miss some informative features that play a key role in recommendations.
\end{itemize}

To sum up, the number of selected feature fields ${K}$ is an essential hyper-parameter of AutoField. A carefully chosen ${K}$ is required to obtain better recommendation results with fewer feature fields.

\subsection{Case Study (RQ5)}
In this subsection, we study whether the selected feature fields of AutoField are optimal from all possible choices on the MovieLens-1M dataset\footnote{ https://grouplens.org/datasets/movielens/1m/}. MovieLens-1M is a benchmark dataset offered by GroupLens Research. There are eight feature fields in this dataset. 
We convert the user ratings on movies (i.e., labels) to binary values. To be specific, user ratings above 3 are set as ``like'' (1), while lower ratings are set as ``dislike'' (0). 
 
\begin{figure}[ht]
    \centering
    \includegraphics[width=0.9\linewidth]{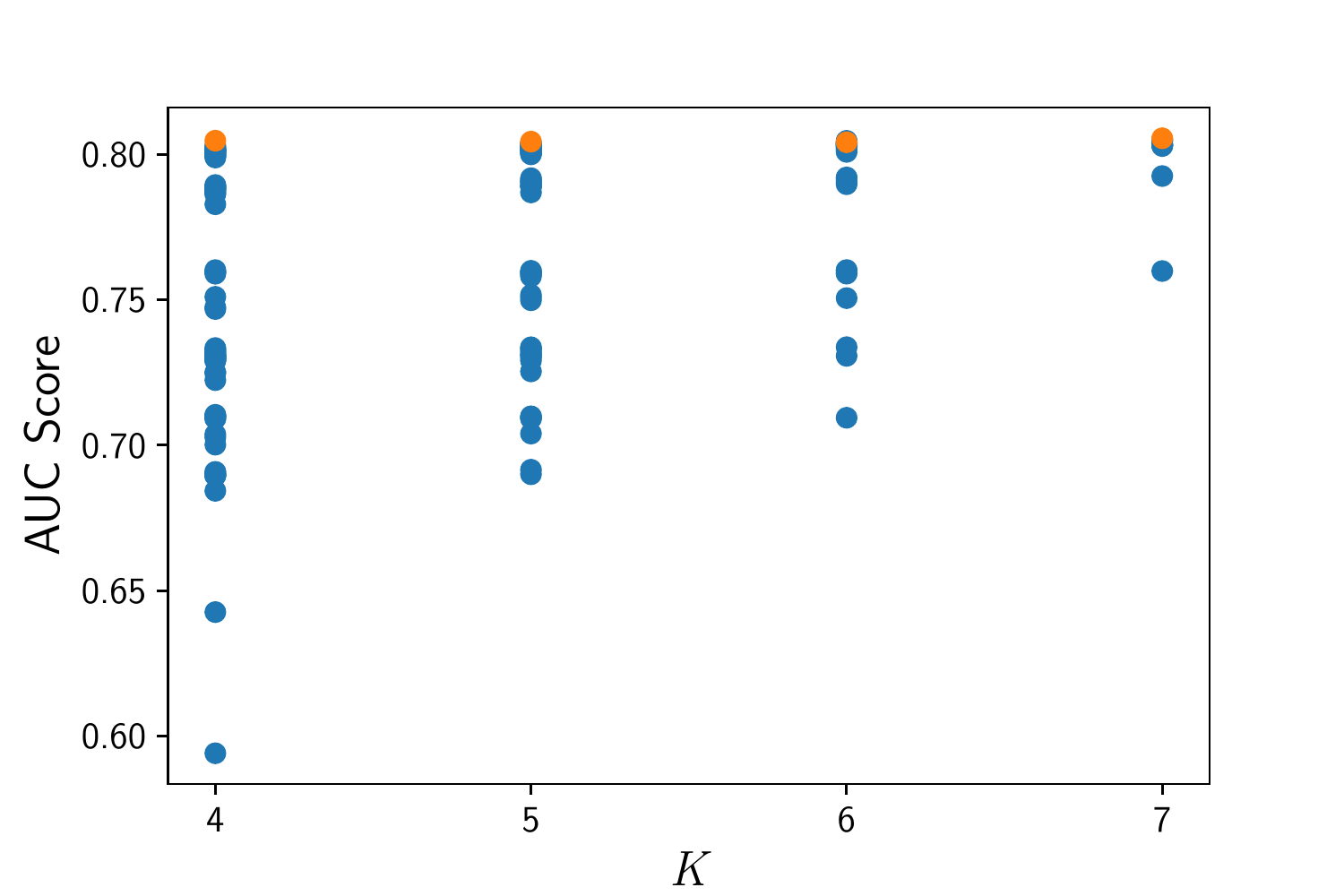}
    \caption{The selection of AutoField (orange) v.s. All possible choices (blue) on MovieLens-1M}
    \label{fig:case}
\end{figure}

The result is visualized in Figure \ref{fig:case}, where the x-axis is the number of selected feature fields ${K}$. Blue points are the AUC scores of all possible feature selection choices, i.e., we enumerate all possible feature subsets. Orange points are the AUC scores of selected feature fields by AutoField with ${K}=4,5,6,7$. We could observe that the selected feature field of AutoField always achieves top performance among all possible solutions.

\section{Related Work} \label{sec:Relat}
In this section, we will discuss the related work. The proposed AutoField is an AutoML-based feature selection model for recommendations. Thus, we will summarize the related work from two perspectives: feature selection techniques and AutoML models.

Feature selection is a popular topic that draws much attention from both academia and industry, which could help to remove redundant or irrelevant feature fields~\cite{JMLR:v22:18-803,liu2019automating,fan2020autofs,zhao2020simplifying,liu2021automated}. In general, feature selection methods could be divided into three categories~\cite{guyon2003introduction}, i.e., the Wrapper, Filter, and Embedded methods. Wrapper methods always elaborate a predictive model as a black-box to evaluate the performance of new subsets of feature fields~\cite{kohavi1997wrappers}. As a result, this kind of solution is computationally intensive. Filter methods are data-oriented, which means the selection process is independent of subsequent models. Consequently, they require less computation and return a general choice~\cite{guyon2003introduction,Yiming}. Embedded methods typically incorporate all feature fields and select features as part of the model construction process. LASSO~\citep{tibshirani1996regression} and Gradient Boosting ~\citep{breiman1997arcing} are representative works in this category. Our framework could be regarded as an Embedded method.

AutoML is the technique that automates the design of machine learning frameworks. In AutoML, the most relevant task to our framework is Neural Architecture Search (NAS), which could be traced back to Zoph et al.~\cite{zoph2017neural}. Zoph et al.~\cite{zoph2017neural} first utilized an RNN controller and designed an encoding method to search for optimal neural network structures with a reinforcement learning optimization strategy. After that, researchers elaborate on new ways to define the search space and devise more efficient optimization strategies to reduce the high training cost of NAS. Pham et al.~\cite{pham2018efficient} design the search space as a directed acyclic graph and come up with a novel training strategy, i.e., weight sharing. Liu et al.~\cite{liu2018darts} devise a differentiable framework DARTS, which motivates our AutoField. There are also some AutoML-based models for feature interaction search~\cite{liu2020autofis,luo2019autocross,song2019autoint}, embedding dimension selection~\cite{zhao2021autoemb,zhao2021autodim,liu2020automated}, and loss function search~\cite{zhao2021autoloss} in recommendations. To the best of our knowledge, we are the first to design an automated feature selection framework for deep recommendation models.
\section{Conclusion}\label{Conclu}

We propose a novel AutoML-based framework, AutoField, to improve the performance of deep recommendation systems and speed up their inference progress by selecting informative feature fields. AutoField is capable of deciding the most effective subset of feature fields from the whole feature space. Specifically, we first provide a basic deep recommendation model to predict users' preferences for items based on the input feature fields. Then, a controller is defined to decide which feature fields should be selected. Then, the feature selection module is introduced to bridge the deep recommendation model and controller. In detail, the feature selection module adapts the $\mathrm{Gumbel{}-{}Softmax}$ operation to simulate the hard selection in the search stage and applies the $\mathrm{Softmax}$ operation in the retraining stage. With optimized AutoField, we finally obtain the optimal subset of feature fields. We carried out substantial experiments to verify the effectiveness of AutoField on two benchmark datasets. The results indicate that AutoField could improve deep recommendation performance with outstanding transferability.

\section*{ACKNOWLEDGEMENT}
This research was partially supported by Start-up Grant (No.9610565) for the New Faculty of the City University of Hong Kong and the CCF-Tencent Open Fund. Also, this research was partially supported by grants from the National Key Research and Development Program of China (Grant No. 2018YFC0117000), and the National Natural Science Foundation of China (Grant No.62072423).

\appendix
\section{Baseline Setting}\label{ap:baselineIntro}

\textbf{Principal Component Analysis(PCA)~\citep{wold1987principal}.} PCA is an eigenvalue decomposition-based multivariate analysis method for dimensionality deduction. 
It projects each data point onto only the first few principal components to obtain lower-dimensional data while preserving as much of the data's variation as possible. 
We choose the number of principal as few as possible, on the condition that the selected components could explain more than $99.9\%$ covariance.  With this assumption, 2 principal components for Avazu and 8 principal components for Criteo are selected. To generate the final results of feature selection, we add up the absolute value of coefficients in all components and pick feature fields with the highest value. The number of final selected feature fields is equal to that of AutoField for a fair comparison.

\textbf{Least Absolute Shrinkage and Selection Operator (LASSO) \citep{tibshirani1996regression}.} LASSO is a regression analysis method for feature selection. By introducing a $L-1$ norm, LASSO forces the sum of squared regression coefficients to be less than a predefined value $\lambda$ and finally sets certain coefficients to zero, achieving the feature selection. It selects the feature fields automatically according to $lambda$. To be specific, we use LASSO implementation in Python Library sci-learn\footnote{https://scikit-learn.org/stable/index.html}. And for fairness, we set $\lambda=0.1$ on Avazu and $\lambda=0.15$ on Criteo. With these settings, similar to AutoField selections, LASSO drops 11 feature fields on Avazu and 5 feature fields on Criteo.

\textbf{Gradient Boosting Technique~\citep{breiman1997arcing}.} Gradient Boosting is a machine learning technique widely used to ensemble models for better performance. Gradient Boosting Regressor(GBR) is an ensemble model based on regression models, while Gradient Boosting Decision Tree(GBDT) is based on decision trees. They both use an impurity-based method to compute feature importance. For GBR, like LASSO, it gives selection directly according to parameters. For GBDT, like PCA, we select the most important $K$ feature fields, the same number as AutoField.

\textbf{IDARTS~\citep{jiang2019improved}.} AutoField framework is encouraged by DARTS\citep{liu2018darts}. IDARTS is another differentiable Neural Architecture Search method, which also could be applied to feature selection problems. The difference is that we use parameter pairs for each feature field that control the feature selection and apply Gumbel-Softmax over each pair to get the final score. IDARTS only assigns one parameter to each feature field and applies softmax over all feature fields together for scoring. 

\bibliographystyle{ACM-Reference-Format}
\bibliography{sample-base}

\end{document}